\documentclass[nofootinbib,aps,11pt]{revtex4-1}
\usepackage{graphicx}
\usepackage{subfigure} 
\usepackage{hyperref}
\usepackage{cancel}
\usepackage{amssymb}
\usepackage{textcomp}
\usepackage{amsmath}
\usepackage{bm}
\usepackage{times}
\usepackage{epsfig}
\usepackage{color}

\newcommand{\eq}{{\rm eq}}
\newcommand{\GeV}{{\rm GeV}}

\begin{document}
\title{\Large Conversion-Driven Dark Matter in $U(1)_{B-L}$}
\bigskip
\author{Jing-Jing Zhang$^1$}
\author{Zhi-Long Han$^1$}
\email{sps\_hanzl@ujn.edu.cn}
\author{Ang Liu$^2$}
\author{Feng-Lan Shao$^2$}
\email{shaofl@mail.sdu.edu.cn}
\affiliation{$^1$School of Physics and Technology, University of Jinan, Jinan, Shandong 250022, China \\
	$^2$School of Physics and Physical Engineering, Qufu Normal University, Qufu, Shandong 273165, China}
\date{\today}
\begin{abstract} 
The new gauge boson $Z'$ in $U(1)_{B-L}$ is widely considered as the mediator of dark matter. In this paper, we propose the conversion-driven dark matter in $U(1)_{B-L}$. The dark sector  contains two Dirac fermions	$\tilde{\chi}_1$ and $\tilde{\chi}_2$ with $U(1)_{B-L}$ charge  0 and $-1$, respectively. A $Z_2$ symmetry is also introduced to ensure the stability of dark matter. The mass term $\delta m \bar{\tilde{\chi}}_1\tilde{\chi}_2$ induces the mixing of dark fermion. Then the lightest dark fermion $\chi_1$ becomes the dark matter candidate, whose coupling to $Z'$ is suppressed by the mixing angle $\theta$. Instead of freezing-out via pair annihilation, we show that the observed relic abundance can be obtained through the conversion processes. We then explore the feasible parameter space of conversion-driven dark matter in $U(1)_{B-L}$. Under various experimental constraints, the conversion-driven dark matter prefers the region with $3\times10^{-6}\lesssim g'\lesssim2\times10^{-4}$ and $0.02~\text{GeV}\lesssim m_{Z'}\lesssim10$~GeV, which is within the reach of future Belle II, FASER and SHiP.
		
\end{abstract}
\maketitle

\section{Introduction}

Various cosmological and astrophysical observations indicate the existence of dark matter (DM) beyond the standard model (SM) of particle physics \cite{Cirelli:2024ssz}. However, the identity of DM is still unknown. Among various candidates, the weakly interacting massive particle (WIMP) through the thermal freeze-out mechanism is one of the most appealing \cite{Arcadi:2017kky,Roszkowski:2017nbc}. To keep the DM in thermal equilibrium at a certain high temperature, sizable interactions between the DM and SM particles are required. A prevalent approach is introducing mediators as a portal to link them, such as the Higgs portal \cite{Patt:2006fw,March-Russell:2008lng,Okada:2010wd,Djouadi:2011aa,Cline:2013gha,Arcadi:2019lka,Arcadi:2021mag} and the vector portal \cite{Alves:2013tqa,DEramo:2016gos,Arcadi:2017hfi,Okada:2018ktp,Blanco:2019hah,Fitzpatrick:2020vba,Liu:2024esf}. Meanwhile, the sizable interactions between the DM and SM particles could also lead to observational signatures at direct detection, indirect detection and collider experiments.

In this paper, we focus on the model of $Z'$ portal DM in $U(1)_{B-L}$. Provided a $Z_2$-odd Dirac fermion $\chi$ with $U(1)_{B-L}$ charge $Q_\chi=-1$ as the DM candidate, the observed DM relic density is obtained via the pair annihilation of $s$-channel $\chi\bar{\chi}\to f\bar{f}$ and $t$-channel $\chi\bar{\chi}\to Z'Z'$ processes, which usually requires the new gauge coupling $g'\sim\mathcal{O}(0.1)$ \cite{Abdallah:2015ter,Klasen:2016qux}. The $Z'$ also mediates the spin-independent DM-nucleon scattering.  As the null result is observed by direct detection experiments \cite{LZ:2022lsv,PandaX-4T:2021bab,DarkSide-50:2023fcw}, the allowed parameter space for Dirac DM is tightly constrained. The direct detection limits might be avoided with Majorana fermion $\chi$ \cite{An:2012va,Escudero:2018fwn}. As a complementary, the direct searches for $Z'$ at colliders have already excluded $m_{Z'}\lesssim5$ TeV with such large gauge coupling \cite{ATLAS:2019erb,CMS:2021ctt}. So the TeV scale DM is naturally expected under the experimental limits. On the other hand, light DM at $Z'$ resonance is also viable  when regarding $Q_\chi$ as a free parameter~\cite{Mohapatra:2019ysk,Nath:2021uqb}. In this way, the gauge coupling $g^\prime$ can be very small to ingratiate various constraints, while appropriate DM coupling $g_\chi$=$Q_{\chi}g^\prime$ can also meet the relic density.

Coscattering \cite{DAgnolo:2017dbv} or Conversion-driven freeze-out \cite{Garny:2017rxs} mechanism is another appealing pathway to accommodate the stringent experimental limits with small DM coupling to SM, where the DM relic density is determined by the inelastic conversions with its heavier dark partner \cite{Garny:2018icg,DAgnolo:2018wcn,Cheng:2018vaj,Junius:2019dci,DAgnolo:2019zkf,Brummer:2019inq,Herms:2021fql,Garny:2021qsr,Filimonova:2022pkj,Heeck:2022rep,Heisig:2024mwr,Heisig:2024xbh,DiazSaez:2024nrq,DiazSaez:2024dzx}. In this paper, we extend the $Z_2$-odd dark sector with two Dirac fermion $\tilde{\chi}_{1,2}$, where the $U(1)_{B-L}$ charges of them are $Q_{\tilde{\chi}_1}=0$ and $Q_{\tilde{\chi}_2}=-1$. A mass term $\delta m \bar{\tilde{\chi}}_1\tilde{\chi}_2$ is further induced, which leads to the mixing of dark fermion. In the mass eigenstate, the lightest dark fermion $\chi_1$ is the DM candidate, whose coupling to $Z'$ is suppressed by the mixing angle $\theta$. So the canonical pair annihilation process is suppressed, while the relic density of $\chi_1$ is determined by the conversion processes.

The structure of this paper is organized as follows. In Sec. \ref{SEC:TM}, we briefly introduce the model of conversion-driven DM in $U(1)_{B-L}$. The calculation of  DM relic density  is discussed in Sec. \ref{SEC:RD}. Then phenomenology of this model is considered in Sec. \ref{SEC:PN}. Finally, we summarize our results in Sec. \ref{SEC:CL}.

\section{The Model}\label{SEC:TM}

In this work, the $U(1)_{B-L}$ model is extended with a $Z_2$-odd dark sector including a Dirac fermion $\tilde{\chi}_1$ and its partner $\tilde{\chi}_2$ beyond the SM. Their $U(1)_{B-L}$ charges are $Q_{\tilde{\chi}_1}=0$ and $Q_{\tilde{\chi}_2}=-1$, respectively. A mass term $\delta m \bar{\tilde{\chi}}_1\tilde{\chi}_2$ then induces the mixing of the dark sector. This term could originate from the interaction of $y \phi \bar{\tilde{\chi}}_1\tilde{\chi}_2$ with $U(1)_{B-L}$ charge $Q_\phi=1$ and the vacuum expectation value $\langle \phi \rangle\neq0$ \cite{Filimonova:2022pkj}. The mass eigenstates are obtained through the rotation as
\begin{equation}
	\begin{pmatrix}
		\chi_1 \\ \chi_2
	\end{pmatrix} =
	\begin{pmatrix}
		\cos\theta & -\sin\theta \\ \sin\theta & \cos\theta
	\end{pmatrix}
	\begin{pmatrix}
		\tilde{\chi}_1 \\ \tilde{\chi}_2
	\end{pmatrix}.
\end{equation}
We assume $m_{\chi_1}<m_{\chi_2}$, thus $\chi_1$ is the DM candidate.
Then in the mass eigenstates, the relevant $Z'$ portal interactions are
\begin{eqnarray}
	\mathcal{L}\supset +Z^\prime_{\mu}g^{\prime}(\cos^2\theta\bar{\chi_2}\gamma^\mu\chi_2-\frac{\sin2\theta}{2}\bar{\chi_2}\gamma^\mu\chi_1-\frac{\sin2\theta}{2}\bar{\chi_1}\gamma^\mu\chi_2+\sin^2\theta\bar{\chi_1}\gamma^\mu\chi_1)-Z^\prime_{\mu}g^{\prime}Q_f\bar{f}\gamma^\mu f
\end{eqnarray}
with $g^\prime$ the $U(1)_{B-L}$ gauge coupling and $Q_f$ the  $U(1)_{B-L}$ charge of SM fermion $f$. 

As we focus on the conversion process, masses of dark fermion are nearly degenerate. It is more convenient to define the fractional mass splitting
\begin{equation}
	\Delta_\chi \equiv \frac{m_{\chi_2}-m_{\chi_1}}{m_{\chi_1}}.
\end{equation}
In general, the phenomenology of conversion-driven DM via $Z'$ portal in $U(1)_{B-L}$ is determined by the parameter set $\{m_{\chi_1},\Delta_\chi,\theta,g',m_{Z'}\}$. However, under various stringent experimental constraints, only the DM at $Z'$ resonance is favored for light DM~\cite{Nath:2021uqb}. In the following studies, we then fix $m_{Z^\prime}/m_{\chi_2}=2.01$ for illustration.

\allowdisplaybreaks

\section{Relic Density}\label{SEC:RD}

Habitually deeming  $Z^\prime$ as a thermal particle, the relevant Boltzmann equations of dark fermions can be expressed as
\begin{eqnarray}\label{Eqn:BE1}
	\frac{dY_{\chi_1}}{dx} &= & -\frac{s}{Hx}\bigg[ \langle \sigma v\rangle_{\chi_1\chi_1\to f\bar{f}}\Big(Y_{\chi_1}^2-(Y_{\chi_1}^{\eq})^2\Big)+\langle \sigma v\rangle_{\chi_1\chi_2\to f\bar{f} }\Big(Y_{\chi_1}Y_{\chi_2}-Y_{\chi_1}^{\eq}Y_{\chi_2}^{\eq}\Big)\nonumber \\
	&-& \langle \sigma v\rangle_{\chi_2 f\to\chi_1 f} \left(Y_{\chi_2}-\frac{Y_{\chi_2}^{\eq}}{Y_{\chi_1}^{\eq}}Y_{\chi_1}Y_f^{\eq}\right)-\langle \sigma v\rangle_{\chi_2\chi_2\to\chi_1\chi_1}\left(Y_{\chi_2}^2-\frac{(Y_{\chi_2}^{\eq})^2}{(Y_{\chi_1}^{\eq})^2}Y_{\chi_1}^2\right)
	\nonumber \\
	&-&\langle \sigma v\rangle_{\chi_1\chi_2\to\chi_1\chi_1}\left(Y_{\chi_1} Y_{\chi_2}-\frac{Y_{\chi_2}^{\eq}}{Y_{\chi_1}^{\eq}}Y_{\chi_1}^2\right)-\langle \sigma v\rangle_{\chi_2\chi_2\to\chi_1\chi_2}\left(Y_{\chi_2}^2-\frac{Y_{\chi_2}^{\eq}}{Y_{\chi_1}^{\eq}}Y_{\chi_1}Y_{\chi_2}\right)
	\nonumber \\
	&-&\frac{\tilde{\Gamma}_{\chi_2\to\chi_1 f\bar{f}}}{s}\left(Y_{\chi_2}-\frac{Y_{\chi_2}^{\eq}}{Y_{\chi_1}^{\eq}}Y_{\chi_1}\right)\bigg]\\	
	\frac{dY_{\chi_2}}{dx} &= & -\frac{s}{Hx}\bigg[ \langle \sigma v\rangle_{\chi_2\chi_2\to f\bar{f}}\Big(Y_{\chi_2}^2-(Y_{\chi_2}^{\eq})^2\Big)+\langle \sigma v\rangle_{\chi_1\chi_2\to f\bar{f}}\Big(Y_{\chi_1}Y_{\chi_2}-Y_{\chi_1}^{\eq}Y_{\chi_2}^{\eq}\Big)\nonumber \\
	&+& \langle \sigma v\rangle_{\chi_2 f\to\chi_1 f}\left(Y_{\chi_2}Y_{f}^{\eq}-\frac{Y_{\chi_2}^{\eq}}{Y_{\chi_1}^{\eq}}Y_{\chi_1}Y_{f}^{\eq}\right)+\langle \sigma v\rangle_{\chi_2\chi_2\to\chi_1\chi_1}\left(Y_{\chi_2}^2-\frac{(Y_{\chi_2}^{\eq})^2}{(Y_{\chi_1}^{\eq})^2}Y_{\chi_1}^2\right)
	\nonumber \\
	&+&\langle \sigma v\rangle_{\chi_1\chi_2\to\chi_1\chi_1}\left(Y_{\chi_1} Y_{\chi_2}-\frac{Y_{\chi_2}^{\eq}}{Y_{\chi_1}^{\eq}}Y_{\chi_1}^2\right)+\langle \sigma v\rangle_{\chi_2\chi_2\to\chi_1\chi_2}\left(Y_{\chi_2}^2-\frac{Y_{\chi_2}^{\eq}}{Y_{\chi_1}^{\eq}}Y_{\chi_1}Y_{\chi_2}\right)
	\nonumber \\
	&+&\frac{\tilde{\Gamma}_{\chi_2\to\chi_1 f\bar{f}}}{s}\left(Y_{\chi_2}-\frac{Y_{\chi_2}^{\eq}}{Y_{\chi_1}^{\eq}}Y_{\chi_1}\right)\bigg]
\end{eqnarray}
where $x=m_{\chi_1}/T$. The entropy density is $s=2\pi^2 g_s(T) T^3/45$. The
Hubble expansion rate is denoted as $H=\sqrt{4\pi^3g_*(T)/45} T^2/m_{pl}$ with the Planck mass $m_{pl}=1.22\times10^{19}~\GeV$. Here,  $g_s(T)$ and $g_\star(T)$ are the number of relativistic degrees of freedom for the entropy and energy density \cite{Husdal:2016haj}, respectively. The thermal average cross sections $\left<\sigma v\right>$ of various channels are calculated numerically by micrOMEGAs~\cite{Belanger:2013oya,Alguero:2022inz,Alguero:2023zol}. 

With the degenerate mass spectrum, the dark partner $\chi_2$ decays into the DM $\chi_1$ via the three-body decays mediated by $Z'$. The corresponding thermal decay width is denoted as
\begin{eqnarray}
\tilde{\Gamma}_{\chi_2\to\chi_1 f\bar{f} }=\frac{\mathcal{K}_1\left(\frac{m_{\chi_2}}{m_{\chi_1}}x\right)}{\mathcal{K}_2\left(\frac{m_{\chi_2}}{m_{\chi_1}}x\right)}\Gamma_{\chi_2\to\chi_1 f\bar{f}}
\end{eqnarray}
where $\mathcal{K}_{1,2}$ are the first and second modified Bessel functions of the second kind, respectively. In the limit of small mass splitting, the three-body decay width can be estimated as \cite{Tsai:2019buq}
\begin{eqnarray}\label{Eq:Chi2Wd}
	\Gamma_{\chi_2\to\chi_1 f\bar{f}}\simeq\frac{{g^\prime}^4\Delta_{\chi}^5m_{\chi_1}^5\sin^2 2\theta}{120\pi^3m_{Z^\prime}^4}.
\end{eqnarray}
In this paper, we use the numerical result of $\Gamma_{\chi_2\to\chi_1 f\bar{f}}$ obtained with micrOMEGAs~\cite{Belanger:2013oya} for more precise calculation. The abundance at thermal  equilibrium can be expressed as~\cite{Alguero:2022inz}
\begin{eqnarray}
	Y_{\chi_1}^{\eq}=\frac{45x^2}{2\pi^4g_s}\mathcal{K}_2(x),~Y_{\chi_2}^{\eq}=\frac{45x^2m_{\chi_2}^2}{2\pi^4g_sm_{\chi_1}^2}\mathcal{K}_2\left(\frac{m_{\chi_2}}{m_{\chi_1}}x\right).
\end{eqnarray}

\begin{figure} 
	\begin{center}
		\includegraphics[width=0.45\linewidth]{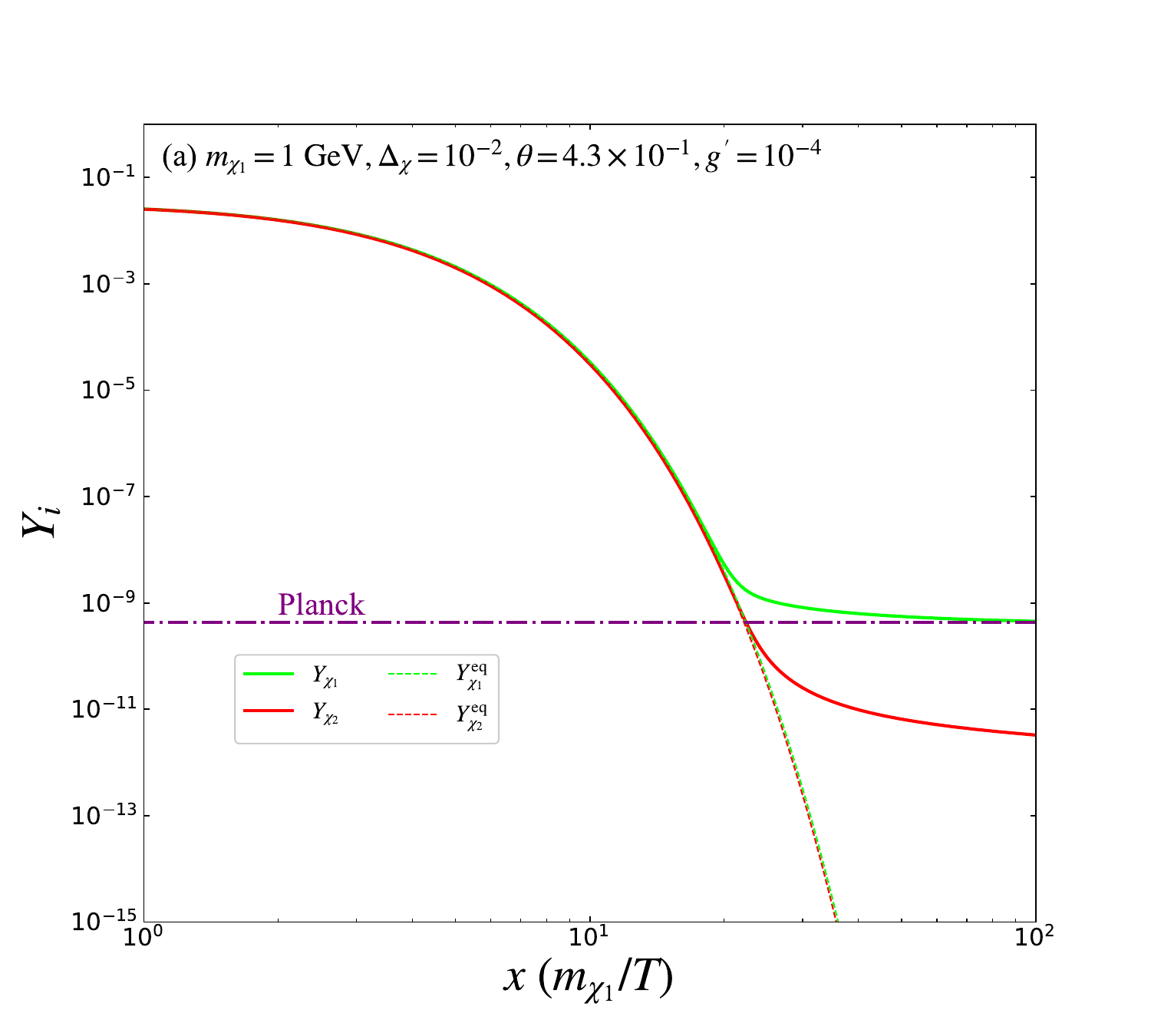}
		\includegraphics[width=0.45\linewidth]{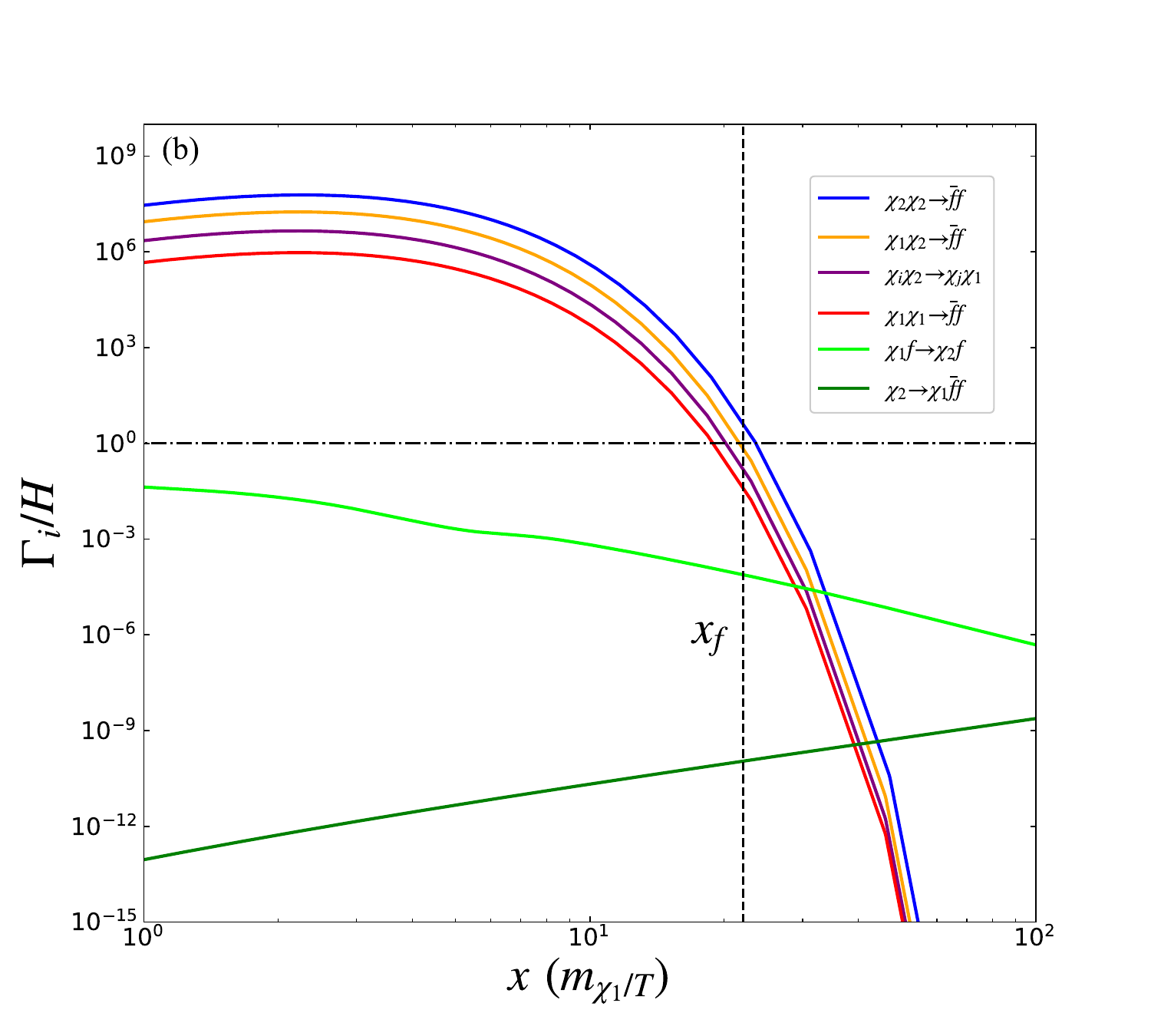}
	\end{center}
	\caption{Panel (a): The evolutions of various abundances $Y_i$. The solid red and green lines represent the abundance evolution of $\chi_1$ and $\chi_2$, while the corresponding dashed lines indicate their respective thermal equilibrium.  The purple dot-dashed line is the Planck observed result for DM with $m_{\chi_1}=1$ GeV \cite{Planck:2018vyg}. Panel (b) : Thermal rates of all relevant interactions.  Here,  $\chi_i \chi_2\to \chi_j \chi_1$ is the sum of the conversion channels $\chi_2\chi_2\to\chi_1\chi_1$, $\chi_1\chi_2\to\chi_1\chi_1$ and $\chi_2\chi_2\to\chi_1\chi_2$. The black vertical dashed line corresponds to the thermal decoupling temperature when $Y_{\chi_1}/Y_{\chi_1}^\eq=2.5$. The horizontal black line is $\Gamma_i=H$. 
 	}
	\label{FIG:fig1}
\end{figure}

A suitable benchmark point, which satisfies the Planck measured DM relic density, is selected to demonstrate the abundance evolution of $\chi_1$ and $\chi_2$ in panel (a) of Fig.~\ref{FIG:fig1}. Meanwhile, the evolution of thermal rates are shown in panel (b) of Fig.~\ref{FIG:fig1}, where the thermal rates of various channels $\Gamma_i$ can be expressed as $n_{a}^{\eq}\left<\sigma v\right>_i$ with $n_a^{\eq}$ the number density of particle $a$ at the thermal equilibrium.  Without controversy, the pair annihilation $\chi_1\chi_1\to f\bar{f}$ causes $\chi_1$ to break away from the thermal equilibrium  too early that $Y_{\chi_1}$ cannot satisfy the observation. Fortunately, the other dark sector processes $\chi_1\chi_2\to f\bar{f}$, $\chi_2\chi_2\to f\bar{f}$ and $\chi_i\chi_2\to\chi_j\chi_1$ still have significant reaction rates, so the conversion from $\chi_2$ to $\chi_1$ can delay the decoupling of $\chi_1$, until the thermal rate of the last related process $\chi_1\chi_2\to f\bar{f}$ is smaller than the Hubble rate $H$. 

Although the evolution of the dark sector in Fig.~\ref{FIG:fig1}  is similar to the well-known coscattering~\cite{DAgnolo:2017dbv}, there are some discrepancies between them under the conditions for coscattering~\cite{DAgnolo:2019zkf}. For instance, coscattering requires that the last reactions to decouple, which change the $\chi_1$ abundance, are the exchange reactions between $\chi_1$ and $\chi_2$, i.e., $\chi_2 f\to\chi_1 f$ or $\chi_2\to\chi_1 f\bar{f}$. Yet it is $\chi_1\chi_2\to f\bar{f}$ in our scenario, meanwhile, the exchange reactions have negligible contributions. With too small mixing $\theta$, the elastic scattering process $\chi_1 f\to\chi_1 f$ might not be able to guarantee $\chi_1$ in kinetic equilibrium with the SM, which could lead to an $\mathcal{O}(10\%)$ distinction on the final relic density compared to the full calculation of unintegrated Boltzmann equation~\cite{Garny:2017rxs}

\begin{figure} 
	\begin{center}
		\includegraphics[width=0.45\linewidth]{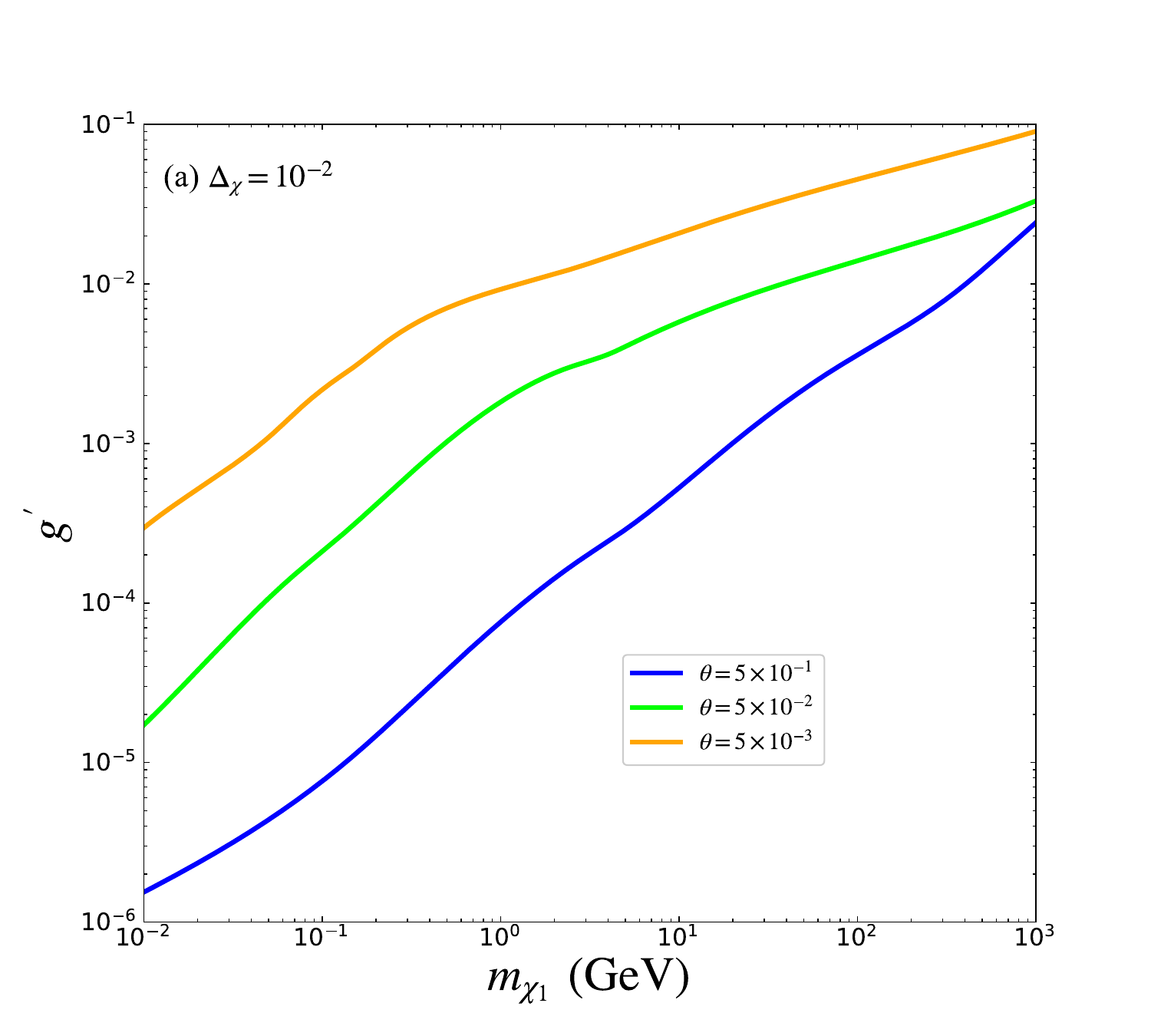}
		\includegraphics[width=0.45\linewidth]{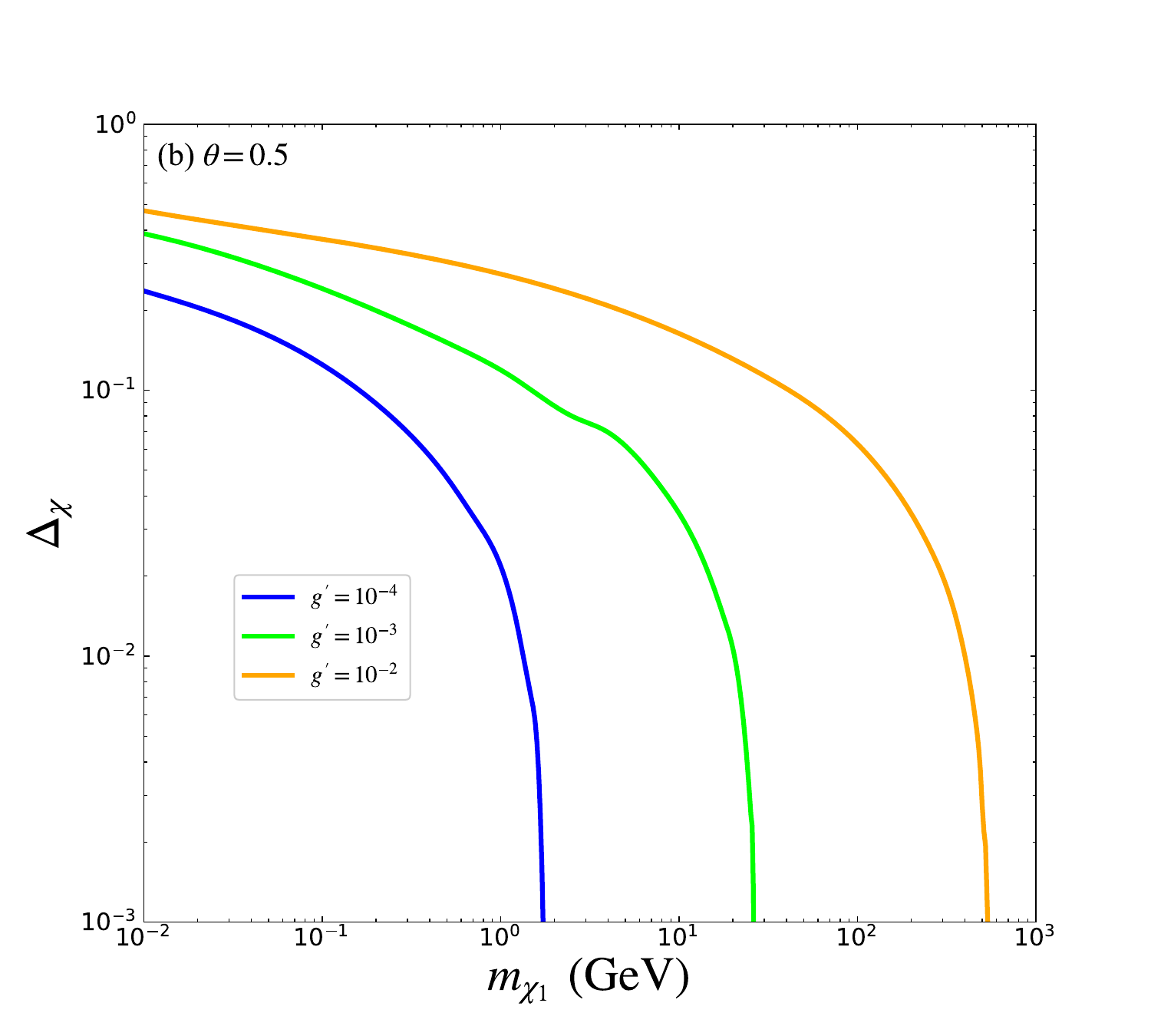}
	\end{center}
	\caption{Viable regions with observed relic abundance for $g^\prime$ as a function of $m_{\chi_1}$ in panel (a), and for $\Delta_{\chi}$ as a function of $m_{\chi_1}$ in panel (b). In panel (a), we fix $\Delta_\chi=10^{-2}$. The blue, green, and orange lines represent $\theta=5\times10^{-1}$, $5\times10^{-2}$ and $5\times10^{-3}$, respectively. In panel (b), $\theta$ is fixed as 0.5, and the three colored curves correspond to $g^\prime=10^{-4}$, $10^{-3}$ and $10^{-2}$.
	}
	\label{FIG:fig2}
\end{figure}

Based on the above benchmark point, we then explore the viable parameter space for correct relic abundance in Fig.~\ref{FIG:fig2}. In panel (a), the fractional mass splitting is fixed as $\Delta_\chi=10^{-2}$ for illustration. The DM relic abundance is determined by $g'$ and $m_{\chi_1}$ for certain mixing $\theta$. Typically, the heavier the DM $\chi_1$ is, the larger the $g'$ is required. Meanwhile, $g'$ will increase as $\theta$ decreases. In this paper, we will no longer consider $\theta>0.5$, since the inhibition of $\theta$ on $\chi_1\chi_1\to f\bar{f}$ will relax for larger $\theta$, which leads to the subdominant contribution of conversion processes. In addition, the conversion-driven mechanism is different from the traditional coannihilation, which usually does not rely on $\theta$. In panel (b), we choose $\theta=0.5$.  By increasing the gauge coupling $g'$, a larger $\Delta_\chi$ is required. However, the fractional mass splitting $\Delta_\chi$ decreases exponentially with the increase of $m_{\chi_1}$ for a specific value of $g'$, which results in an upper bound on $m_{\chi_1}$ when $\Delta_\chi\lesssim10^{-2}$.

\section{Phenomenology}\label{SEC:PN}

\subsection{Constraints on $Z^\prime$}

\begin{figure} 
	\begin{center}
		\includegraphics[width=0.45\linewidth]{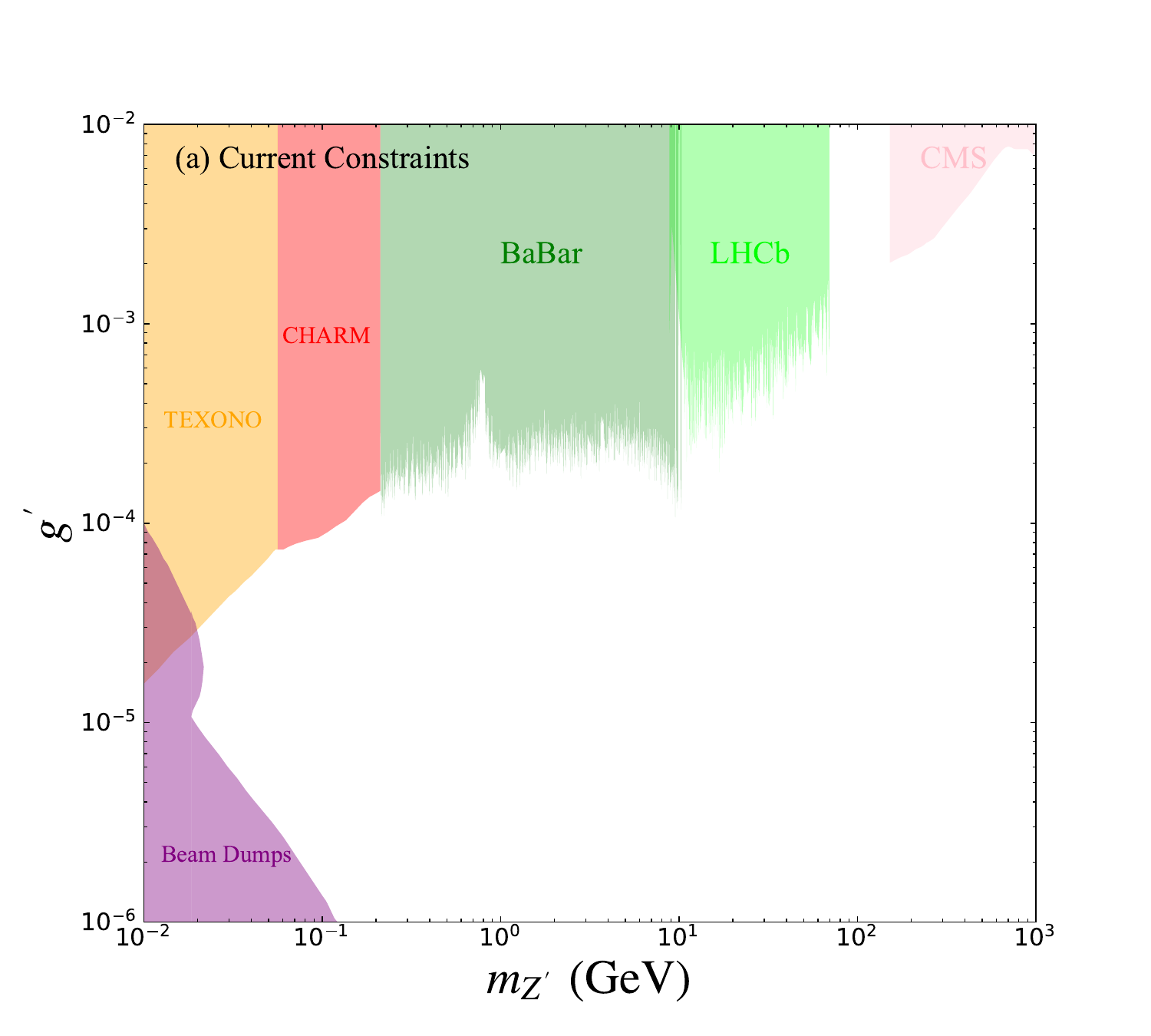}
		\includegraphics[width=0.45\linewidth]{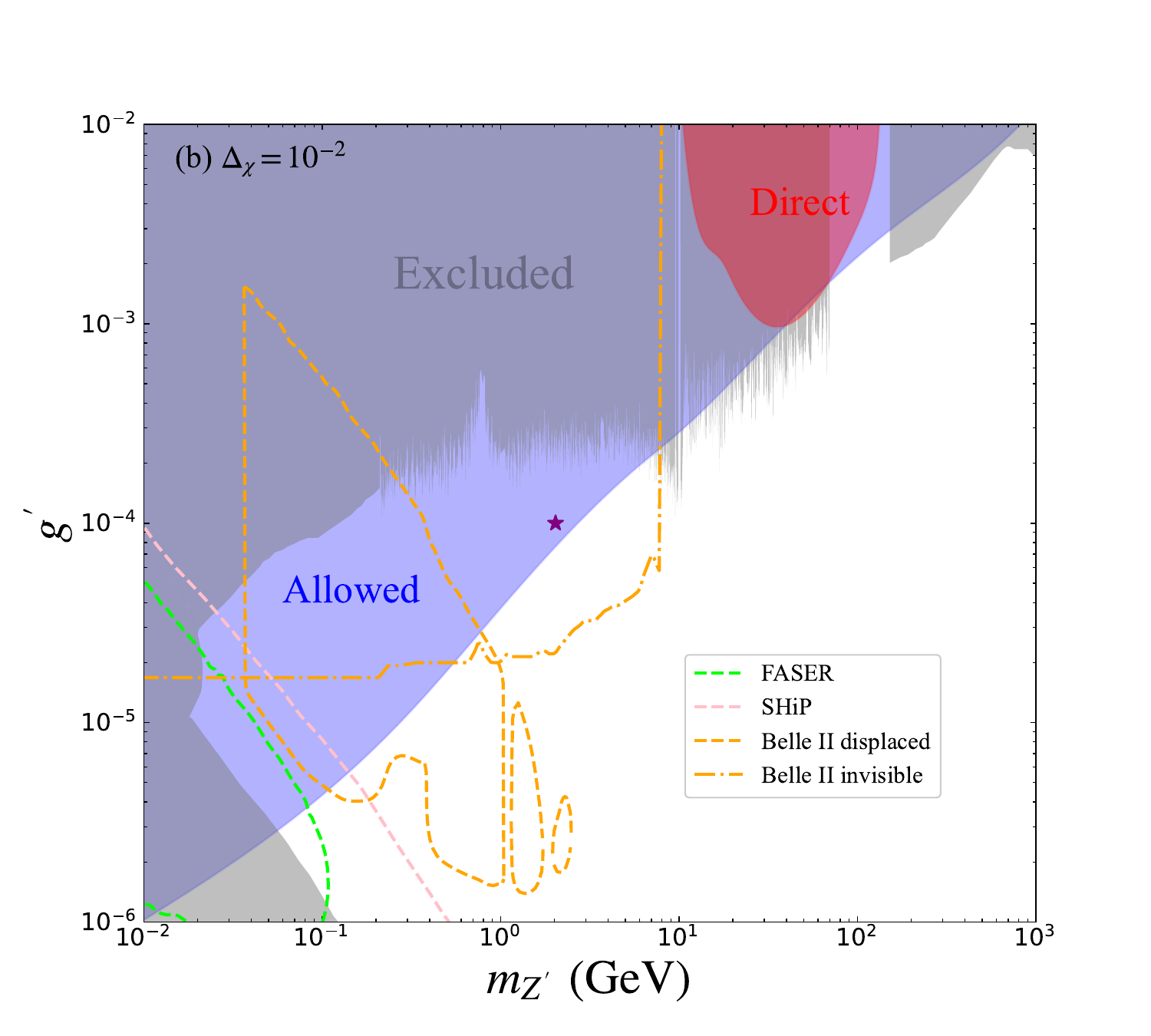}
	\end{center}
	\caption{Panel (a): Various constraints on the $Z'$ in $U(1)_{B-L}$. Panel (b): Allowed parameter space for conversion-driven DM. The blue region represents the allowed parameter space for conversion-driven DM with $\theta<0.5$. The purple star corresponds to the benchmark point in Fig.~\ref{FIG:fig1}. The gray areas have been excluded by the current experimental search for $Z'$. The red shadow  is excluded by the current direct detection experiments LZ and PandaX-4T~\cite{LZ:2022lsv,PandaX-4T:2021bab}.  The orange dashed and dotdashed lines correspond to the future Belle II sensitivity for long-lived~\cite{Ferber:2022ewf} and invisible~\cite{Dolan:2017osp} $Z^\prime$, respectively. Meanwhile, the green and pink dashed lines stand for the expected sensitivities of long–lived $Z^\prime$ searching at FASER~\cite{FASER:2018eoc} and 
    SHiP~\cite{Alekhin:2015byh}.  }
	\label{FIG:fig3}
\end{figure}

In the above conversion-driven scenario, the gauge boson is also relatively light as we assume $m_{Z'}=2.01m_{\chi_2}\simeq 2 m_{\chi_1}$. With direct couplings to SM fermion of $Z'$ in $U(1)_{B-L}$, the sub-GeV light gauge boson can be tested at various beam dump \cite{CHARM:1985anb,Bauer:2018onh,Dev:2021qjj} and neutrino scattering experiments \cite{TEXONO:2009knm,Bilmis:2015lja,KA:2023dyz}. For GeV scale $Z'$, tight constraint comes from the search for dark photon at BaBar \cite{BaBar:2014zli,BaBar:2017tiz} and LHCb \cite{LHCb:2017trq,LHCb:2019vmc}. Meanwhile, the search for gauge boson in the dilepton channel could probe TeV scale $Z'$ at LHC~\cite{CMS:2021ctt,ATLAS:2019erb}. In panel (a) of Fig~\ref{FIG:fig3}, we summarize the current experimental constraints on $Z'$. 

The conversion-driven scenario is viable in the parameter space with $3\times10^{-6}\lesssim g'\lesssim2\times10^{-4}$ and $0.02~\text{GeV}\lesssim m_{Z'}\lesssim10$~GeV. As shown in Fig.~\ref{FIG:fig2}, the allowed region is expanded as $\theta$ decreases, which is roughly allowed within the range of $\theta\in[0.05,0.5]$ under current constraints. The viable area is obtained with $\Delta_\chi=10^{-2}$. Although decreasing $\Delta_{\chi}$ will not alert the allowed region, increasing $\Delta_{\chi}$ could significantly diminish $m_{Z'}$. The benchmark point presented in Fig.~\ref{FIG:fig1} is promising to be detected by future Belle II~\cite{Ferber:2022ewf,Dolan:2017osp}, which is sensitive to $m_{Z^\prime}\lesssim8$~GeV and $g^\prime\gtrsim4.4\times10^{-6}$. Furthermore, the FASER~\cite{FASER:2018eoc} and SHiP~\cite{Alekhin:2015byh} experiments have more potential for smaller $g^\prime$ at sub-GeV. It is obvious that the conversion-driven region is fully in the reach of future experiments.

\subsection{Constraints on Dark Matter $\chi_1$}

\begin{figure} 
	\begin{center}
		\includegraphics[width=0.45\linewidth]{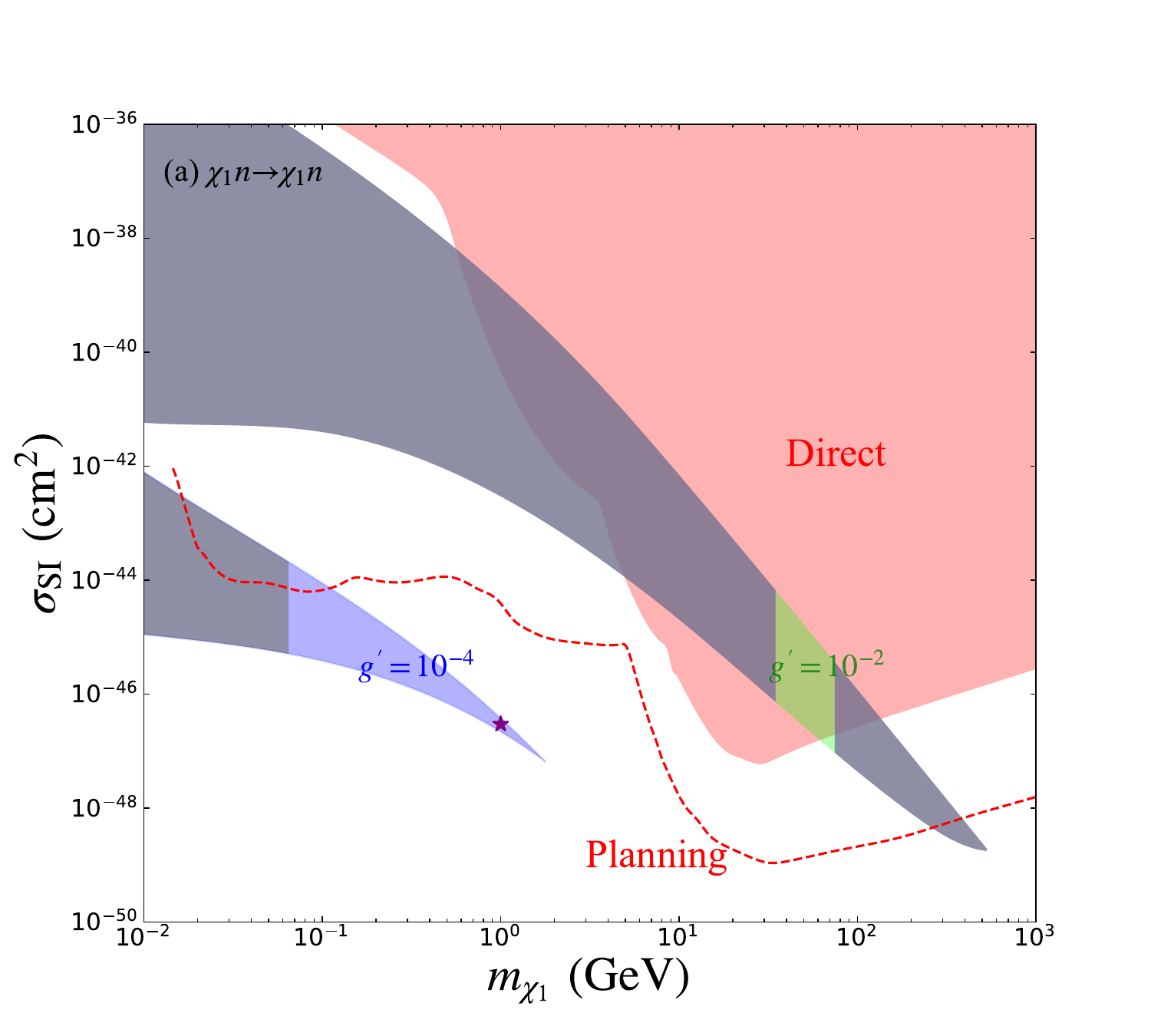}
		\includegraphics[width=0.45\linewidth]{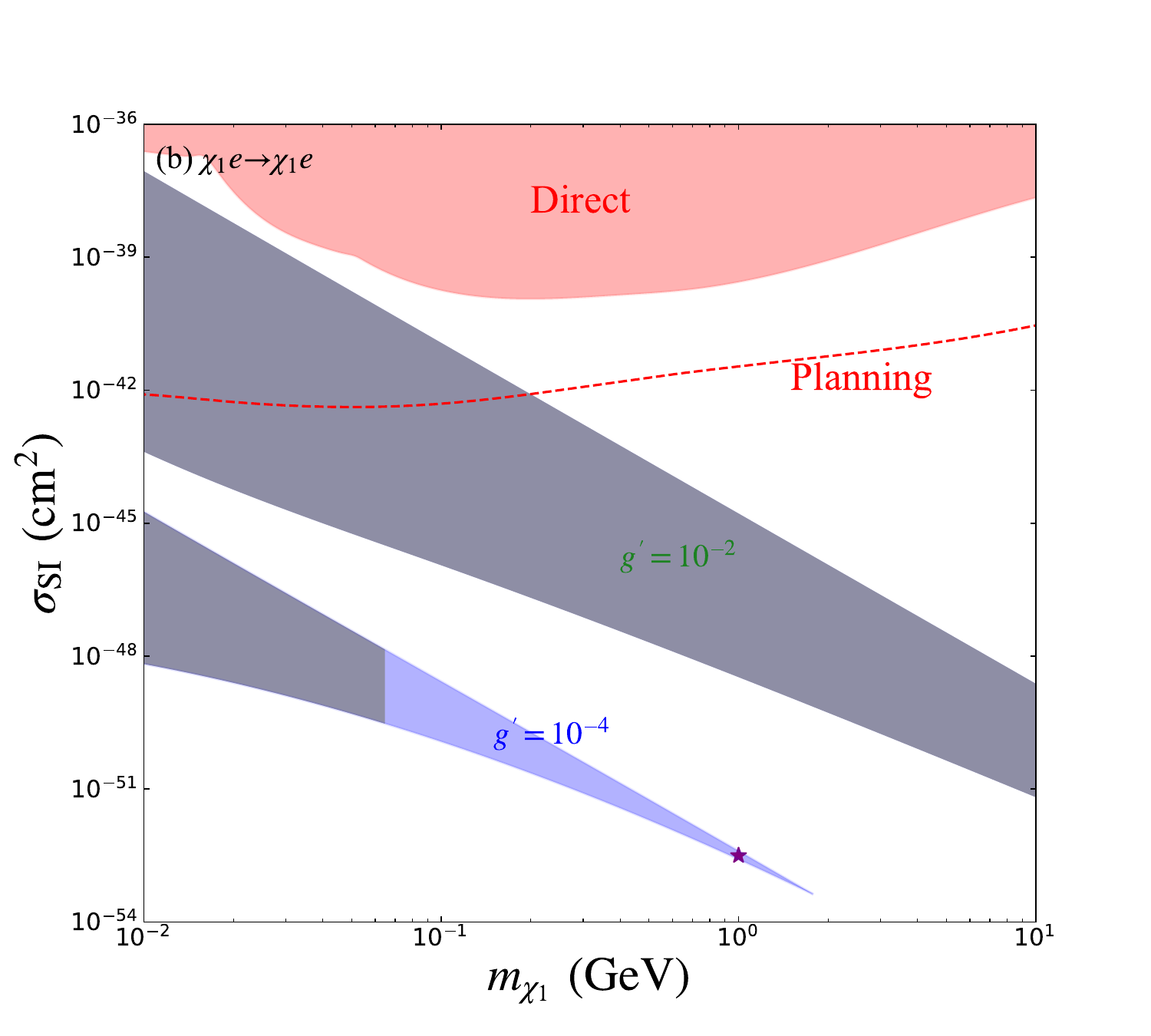}
	\end{center}
	\caption{Panel (a): The spin-independent DM-nucleon scattering cross section. The green and blue regions correspond to $g'=10^{-2}$ and $g'=10^{-4}$, respectively. The gray region is already excluded by collider searches of $Z'$. The red region is excluded by the current direct detection experiments from LZ \cite{LZ:2022lsv}, PandaX-4T \cite{PandaX-4T:2021bab}, and DarkSide-50 \cite{DarkSide-50:2023fcw}. The red dashed line is the sensitivity of future experiments from DarkSide-LowMass \cite{GlobalArgonDarkMatter:2022ppc}, SuperCDMS \cite{SuperCDMS:2016wui}, and LZ \cite{LZ:2015kxe}.
	Panel (b): Same as panel (a), but for DM-electron scattering. The red region is excluded by the current experiment from XENON1T~\cite{XENON:2019gfn,XENON:2021qze}, DarkSide-50~\cite{DarkSide:2022knj}, and DAMIC-M~\cite{DAMIC-M:2023gxo}. The red dashed line is the future sensitivity of the BRN experiment \cite{Essig:2022dfa}. }
	\label{FIG:fig4}
\end{figure}

The canonical $Z'$ portal Dirac DM in $U(1)_{B-L}$ usually suffers stringent constraints from direct detection experiments. In the conversion-driven scenario, the DM $\chi_1$ can scatter with nucleon through mediator $Z^\prime$. The spin-independent scattering cross section is:
\begin{eqnarray}\label{Eqn:dd}
	\sigma_{\rm SI}=\frac{m_{\chi_1}^2 m_n^2 \sin^2 \theta {g^\prime}^4}{\pi m_{Z^\prime}^4(m_{\chi_1}+m_n)^2},
\end{eqnarray}
where $n$ stands for nucleons, and $m_n\simeq0.939$ GeV. Therefore, the cross section of conversion-driven DM is further suppressed by the mixing $\theta$. 

By requiring correct relic abundance, it is already shown in Fig.~\ref{FIG:fig2} that the required value of $\theta$  is fixed for certain $g'$ and $\Delta_{\chi}$. In panel (a) of Fig.~\ref{FIG:fig4}, we show the predicted scattering cross section for $g'=10^{-2}$ and $g'=10^{-4}$ while varying $\Delta_{\chi}$ in the range of $[0.001,0.5]$. For $g'=10^{-2}$,  we found that the current direct detection restrictions could exclude the parameter space with $m_{Z^\prime}\in[10,150]$ GeV. The direct detection exclusion region is depicted as the red area in panel (b) of Fig.~\ref{FIG:fig3}, which has excluded $g'\gtrsim10^{-3}$ with $m_{Z'}\sim\mathcal{O}(10)$ GeV. This excluded region covers the collider insensitive part near the SM $Z$ resonance. Meanwhile, the scattering cross section for $g'=10^{-4}$ is typically less $10^{-44}~\text{cm}^2$ with sub-GeV DM. This region is within the neutrino fog \cite{Akerib:2022ort}, thus it is hard to be detected at even future nucleon scattering experiments.

For $m_{\chi_1}$  below GeV, electron recoils present a more promising outlook~\cite{Knapen:2017xzo,Ema:2018bih}. With a heavy mediator in this paper, it is sufficient to replace $m_n$ with $m_e$ in Eqn.~\eqref{Eqn:dd} to obtain the DM-electron scattering cross section. The results are shown in panel (b) of Fig.~\ref{FIG:fig4}. For $g'=10^{-4}$. The predicted value of $\sigma(\chi_1 e\to \chi_1e)$ is less than $10^{-47}~\text{cm}^2$, which is also beyond the reach of future experiments.

As masses of dark fermion nearly degenerate, the inelastic up-scattering $\chi_1n\to\chi_2n$ could lead to an observable signature when the mass splitting $\delta_\chi=m_{\chi_2}-m_{\chi_1}$ is less than $\mathcal{O}(100)$ keV for DM above the GeV scale \cite{Tucker-Smith:2001myb}. According to Eqn.~\eqref{Eq:Chi2Wd}, a tiny fractional mass splitting may lead lifetime of $\chi_2$ longer than our universe, so we consider the fractional mass splitting $\Delta_\chi>10^{-3}$ in this paper. In the conversion-driven scenario, we then have $\delta_\chi = m_{\chi_1} \Delta_{\chi}\gtrsim 10$ keV if $\Delta_{\chi}\gtrsim10^{-3}$, which is not in the accessible region of up-scattering for $m_{\chi_1}\lesssim1$ GeV \cite{CarrilloGonzalez:2021lxm}. 

It is also worth mentioning that the dark matter could be accelerated by the inelastic scattering off cosmic proton $\chi_1 p \to \chi_2 p$ \cite{Bringmann:2018cvk}. The relativistic component of dark matter from cascade decay $\chi_2\to \chi_1 f\bar{f}$ would pass the energy threshold of up-scattering $\chi_1n\to\chi_2n$ ~\cite{Bell:2021xff}. If the dark partner $\chi_2$ is long-lived, the down-scattering processes $\chi_2n\to\chi_1n$ and $\chi_2e\to\chi_1e$ also lead to observable signatures \cite{Graham:2010ca,Aboubrahim:2020iwb}. The current experiments are sensitive to the region with $g'\gtrsim10^{-2}$ \cite{Bell:2021xff,Aboubrahim:2020iwb}, which is far beyond the viable region of conversion-driven scenario in this paper.

For sub-GeV DM, another tight constraint comes from the Cosmic Microwave Background (CMB), due to additional energy injection into the SM plasma by DM annihilation during recombination. The Planck experiment has constrained the effective parameter \cite{Planck:2018vyg}
\begin{equation}
	p_\text{ann}=\frac{1}{2} f_\text{eff} \frac{\langle \sigma v \rangle_\text{CMB}}{m_{\chi_1}} < 3.2\times 10^{-28} \text{cm}^3 \text{s}^{-1} \text{GeV}^{-1},
\end{equation}
where $f_\text{eff}$ is the efficiency of energy transfer to SM plasma. For conversion-driven DM in this paper, the annihilation cross section is suppressed by the mixing angle $\theta$ as well as the off-resonance effect \cite{Bernreuther:2020koj}, as we consider $m_{Z'}\gtrsim2 m_{\chi_1}$. The typical annihilation cross section of DM $\chi_1$ during recombination is less than $10^{-32}~\text{cm}^3\text{s}^{-1}$, which is under the CMB limit for sub-GeV DM \cite{Brahma:2023psr}. 

\subsection{Constraints on Long-lived $\chi_2$}

With relatively small mass splitting, the three-body decay of $\chi_2\to\chi_1\bar{f}f$ through off-shell $Z^\prime$ is the dominant channel of dark fermion $\chi_2$, thus $\chi_2$ is usually long-lived. For sub-GeV $\chi_2$, the most appealing channel is $\chi_2\to \chi_1 e^+e^-$, meanwhile, the only viable channel is $\chi_2\to\chi_1 \bar{\nu}\nu$ if the mass splitting $\delta_\chi=m_{\chi_2}-m_{\chi_1}<2m_e$ \cite{Foguel:2024lca}. Once allowed, the visible decay $\chi_2\to\chi_1 e^+e^-$ in principle would lead to multilepton signature at colliders via processes as $pp\to Z'\to \chi_2 \bar{\chi}_2 \to\chi_1\bar{\chi_1}e^+e^-e^+e^-$, which is different from the dilepton signature of inelastic DM as $pp\to Z'\to \chi_1 \chi_2 \to \chi_1 \chi_1 \ell^+\ell^-$ \cite{CMS:2023bay}. 

\begin{figure} 
	\begin{center}
		\includegraphics[width=0.45\linewidth]{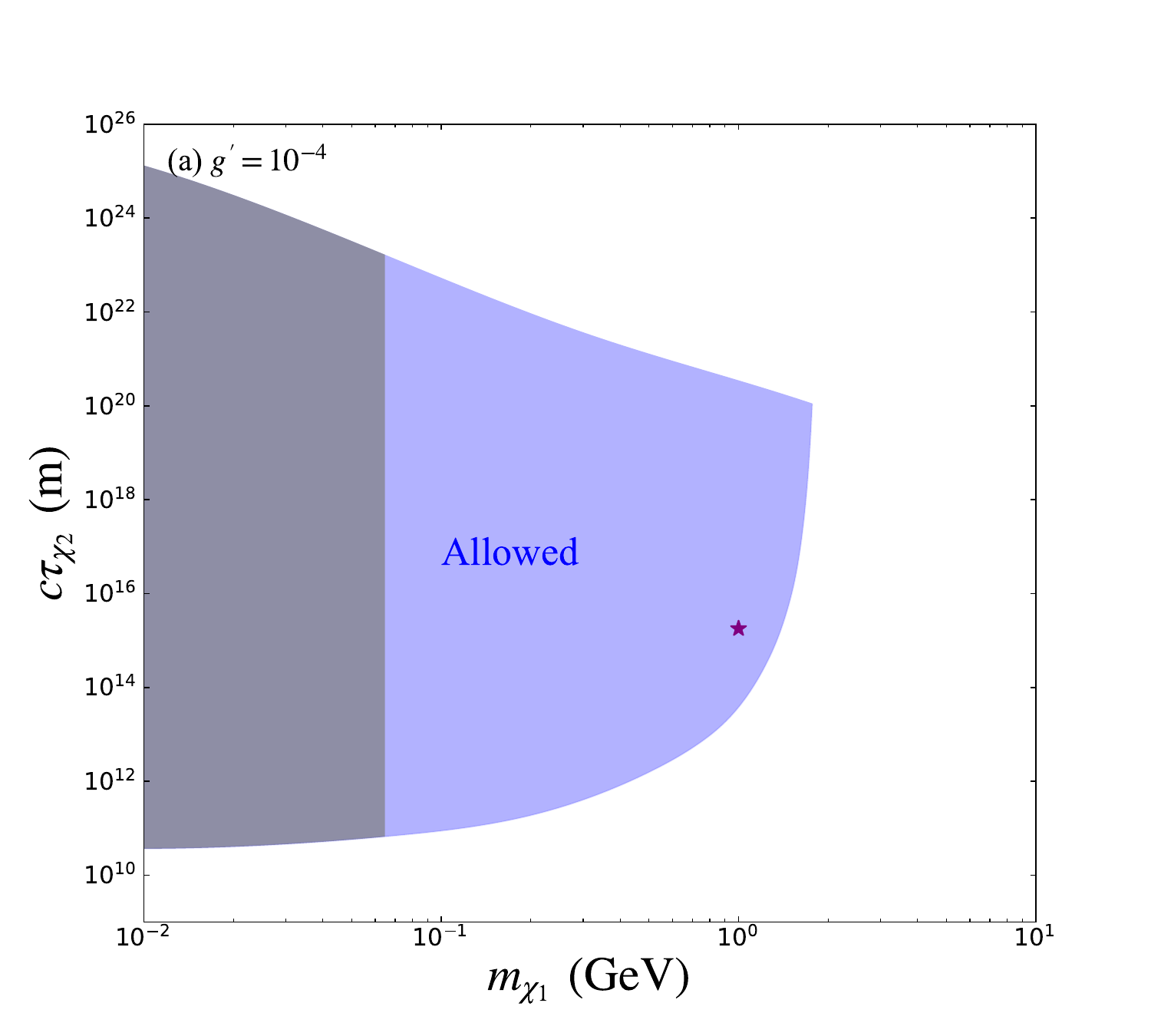}
		\includegraphics[width=0.45\linewidth]{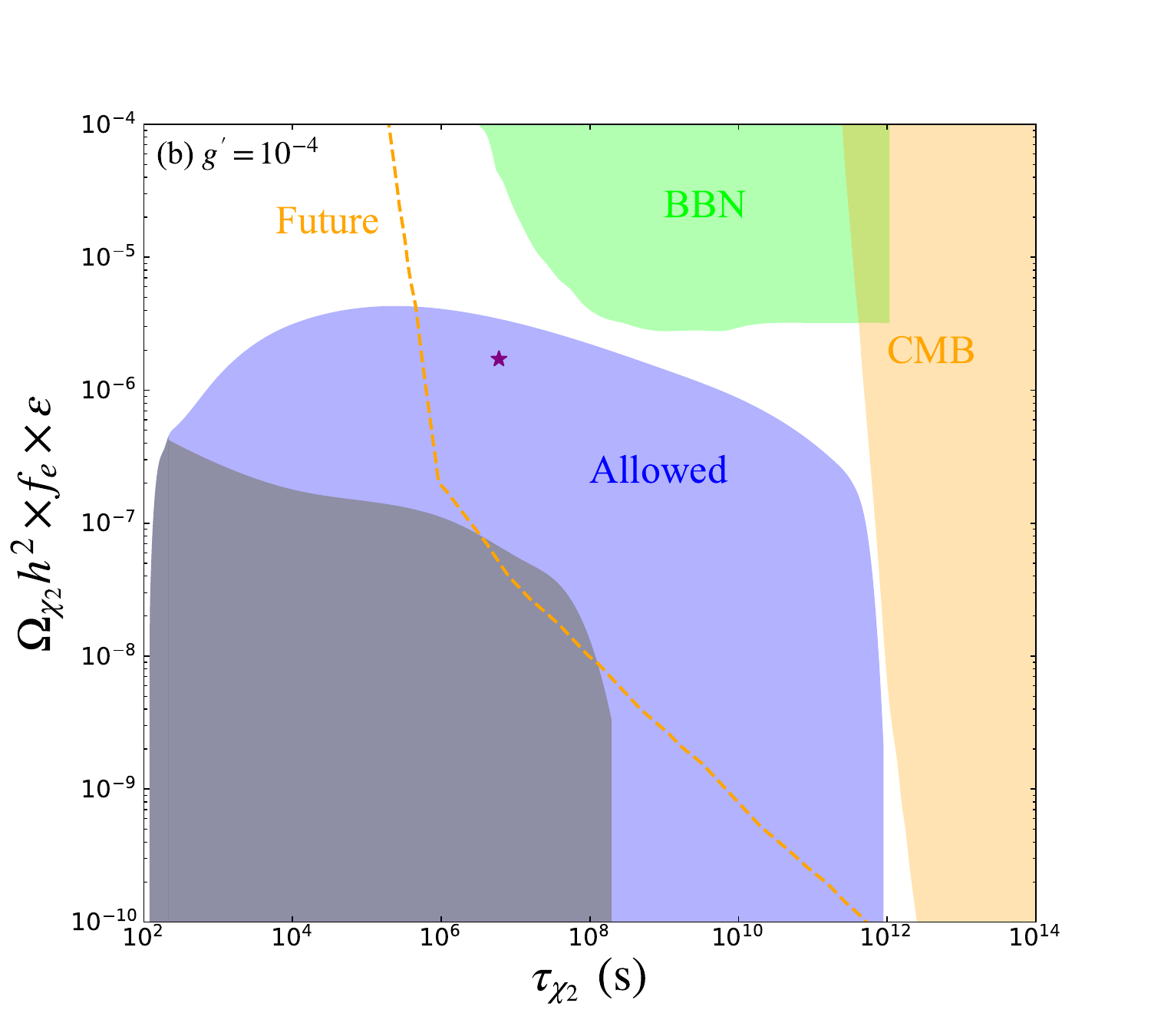}
	\end{center}
	\caption{Panel (a): Decay length of $\chi_2$ in the allowed parameter space. The gray and dark blue areas represent the same as that in (b) in Fig.~\ref{FIG:fig3}. 
	Panel (b): CMB and BBN constraints on long-lived $\chi_2$. The orange and green areas are excluded by CMB ~\cite{Acharya:2019uba} and BBN  \cite{Kawasaki:2017bqm}. The orange dashed line represents the future CMB-S4 results ~\cite{Lucca:2019rxf}.
	}
	\label{FIG:fig5}
\end{figure}

The theoretical decay length of $\chi_2$ is shown in panel (a) of Fig.~\ref{FIG:fig5}. Due to relatively small gauge coupling $g'$ and fractional mass splitting $\Delta_{\chi}$, we find that the decay length $c\tau_{\chi_2}$ is typically larger than $10^{10}$ m, which is far from the collider sensitivity region \cite{Berlin:2018jbm}. So $\chi_2$ is also invisible at colliders. The observable signal would be monophoton as $e^+e^-\to Z' \gamma \to \chi_2\bar{\chi}_2  \gamma$. However, the dominant invisible decay of gauge boson is $Z'\to \nu\bar{\nu}$ in this paper for $Q_{\tilde{\chi}_2}=-1$, while the decay width of $Z'\to \chi_2 \bar{\chi}_2$ is suppressed by the phase space as $m_{Z'}\simeq 2 m_{\chi_2}$. A precise measurement of invisible $Z'$ decay is required to extract the contribution of $\chi_2$.

Furthermore, the energy injection from the delayed decay of $\chi_2$  would affect the observations of cosmic microwave background (CMB) and big-bang nucleosynthesis (BBN) in the early universe. Assuming $\chi_2\to \chi_1 e^+e^-$ being allowed, we illustrate the impact of cosmological constraints on the allowed regions in panel (b) in Fig.~\ref{FIG:fig5}. The current BBN bound imposes strict constraint on $\tau_{\chi_2}\lesssim10^{12}$ s and $\Omega_{\chi_2}h^2f_e\epsilon\gtrsim3\times10^{-6}$. Longer $\tau_{\chi_2}$ will be excluded by the present CMB bound.  These cosmological constraints are derived under the assumption that $\chi_2$ decays solely into $e^+e^-$.  In order to accommodate these constraints, the relic density of $\chi_2$ should multiply by $f_e$ the branch ratio of $\chi_2\to\chi_1e^+e^-$ and $\epsilon =(m_{\chi_2}^2-m_{\chi_1}^2)/2m_{\chi_2}^2$ the fraction of the energy of $\chi_2$ that has been transferred to electron. It is evident that the viable region is under the current limits. The future CMB-S4 will significantly limit the allowed region with $\tau_{\chi_2}\gtrsim 4.5\times10^5$ s.

We should mention that $\tau_{\chi_2}>10^{12}$ s is possible. However, such a long lifetime needs the mass splitting $\delta_\chi<2 m_e$. So $\chi_2\to\chi_1 \bar{\nu}\nu$ is the only decay mode. For the sub-GeV scale $\chi_2$, the CMB observation set no limit due to too small fraction of energy injection into SM plasma from neutrinos ~\cite{Hambye:2021moy}. The neutrinos from delayed decay will contribute to the effective number of relativistic neutrino species $N_{\rm eff}$ ~\cite{Liu:2022cct}. The allowed parameter space predicts  $(f_\nu\epsilon~\Omega_{\chi_2}/\Omega_{\chi_1})^2\tau_{\chi_2}\sim\mathcal{O}(10^7)$ s, which is far below the current Planck limit $(f_\nu\epsilon~\Omega_{\chi_2}/\Omega_{\chi_1})^2\tau_{\chi_2}\lesssim5\times10^{9}$ s ~\cite{Hambye:2021moy}. Therefore, the conversion-driven DM with too small mass splitting $\delta_\chi<2m_e$ is hard to probe through cosmological observations.

The relic abundance of light DM might be affected when the freeze-out temperature $T_f$ is in proximity to the QCD phase transition temperature $T_{QCD}\sim 0.2$ GeV~\cite{Steigman:2012nb}. With sub-GeV DM in the viable region, the freeze-out temperature $T_f=m_{\chi_1}/x_f<1/25 ~\text{GeV}<T_{QCD}$, hence the effects of the QCD phase transition on relic abundance can be neglected \cite{Drees:2015exa}. The dark particles generated at the supernova will carry energy away, which could speed up the supernova's cooling.   The supernova 1987A is sensitive to the scope of feeble coupling $g^\prime\in[10^{-10},10^{-7}]$ \cite{Chang:2018rso}, which is below the viable region in Fig.~\ref{FIG:fig3}.

\section{Conclusion}\label{SEC:CL}

In this work, we extend the dark sector with two Dirac fermion $\tilde{\chi}_{1,2}$, which carry  $U(1)_{B-L}$ charges  0 and $-1$, respectively. We impose a $Z_2$ symmetry to ensure the stability of dark matter. A mass term $\delta m \bar{\tilde{\chi}}_1\tilde{\chi}_2$ then induces the mixing of the dark sector. In the mass eigenstate, the lightest dark fermion $\chi_1$ is the dark matter candidate. Different from the conventional models, the couplings of dark matter to gauge boson in this paper are determined by the mixing angle $\theta$ between the dark fermions. For a small mixing angle $\theta$, the pair annihilation of dark matter is suppressed. The correct relic density of $\chi_1$ is obtained via the conversion processes with a nearly degenerate mass spectrum of dark fermion $m_{\chi_2}\simeq m_{\chi_1}$.

We then explore the viable parameter space of the conversion-driven dark matter, which is  $3\times10^{-6}\lesssim g'\lesssim2\times10^{-4}$ and $0.02~\text{GeV}\lesssim m_{Z'}\lesssim10$~GeV under various constraints. It should be noted that such region is obtained through the $Z'$ resonance with $m_{Z'}\sim 2 m_{\chi_1}$, while the mixing angle is limited within the range $[0.05,0.5]$. The direct search of  $Z^\prime$ at the current colliders excludes most of the parameter space above $g^\prime\sim10^{-4}$. Meanwhile, the  allowed areas are expected to be detected by future FASER,  SHiP and Belle II experiments.

For the light conversion-driven dark matter, we find that direct detection experiments, including DM-nucleon scattering, DM-electron scattering, and inelastic up-scattering, are all hard to have positive signatures. The annihilation cross section is also suppressed by the off-resonance effect, which leads to the dark matter unpromising at indirect detection experiments. However, the dark partner $\chi_2$ could induce an observable signature when the mass splitting $m_{\chi_2}-m_{\chi_1}>2m_e$. The energy injection from the delayed decay of $\chi_2\to \chi_1 e^+e^-$ is within the reach of future CMB-S4 experiment. 

\section*{Acknowledgments}

This work is supported by the National Natural Science Foundation of China under Grant No. 12375074, 12175115, and 11805081, Natural Science Foundation of Shandong Province under Grant No. ZR2019QA021.


\end{document}